\documentstyle[prl,aps,multicol]{revtex}
\input{epsf.sty}

\begin{document}

\title{On the validity of the Aharonov-Bergmann-Lebowitz rule}

\author{ Lev Vaidman} 

\address{School of Physics and Astronomy, 
Raymond and Beverly Sackler Faculty of Exact Sciences, \\
Tel-Aviv University, Tel-Aviv 69978, Israel.} 

\date{}

\maketitle

\begin{abstract}
  It is argued that the proof of Cohen [Phys. Rev. {\bf A 51}, 4373
  (1995)] which shows that  an application of the
  Aharonov-Bergmann-Lebowitz (ABL) rule leads to contradiction with
  predictions of quantum theory is erroneous. A generalization of the
  ABL rule for the case of an incomplete final measurement (which is
  needed for the analysis of Cohen's proof) is presented.
\end{abstract}


\begin{multicols}{2}
  
  Cohen\cite{Co} examines a few surprising results that have been
  obtained for pre- and post-selected quantum system by applying the
  Aharonov, Bergmann and Lebowitz (ABL) rule\cite{ABL}. Following
  Sharp and Shanks\cite{SS} he proves that the ABL rule is not valid
  in general by showing that in a particular situation it leads to
  a prediction contradicting  quantum theory. He
  claimed, however, that the ABL rule is valid for a special class of
  situations which correspond to ``consistent histories''\cite{G}.
  This limitation, if true, reduces significantly the importance of
  the ABL rule.  In this
  comment I will argue that there is a crucial error in Cohen's proof
  of the inconsistency of the ABL rule with  quantum theory.

The  proof of Cohen  is a variation
of the proof given earlier by Sharp  and Shanks\cite{SS}.
I have showed in details in Ref.\cite{V-coun} the flaw in these
arguments, and  here I will only present the key point and discuss
details which are specific for Cohen's proof.

Cohen considers a modified Mach-Zehnder apparatus with a possible
measurement performed by a ``which way'' detector  $D_3$ (that lets detected
particles  pass through), see Fig. 1.
 Then he applies the ABL rule in a counterfactual sense, and arrives at
a 
contradiction with  quantum theory. This leads him to
reject  the ABL rule for counterfactual situations.
I argue that the contradiction  obtained by Cohen  follows from a logical
error in his equation and not from an
inapplicability of the ABL formula in this case. 
\vskip .5cm
\epsfysize=5.3 cm
 \leftline{\epsfbox{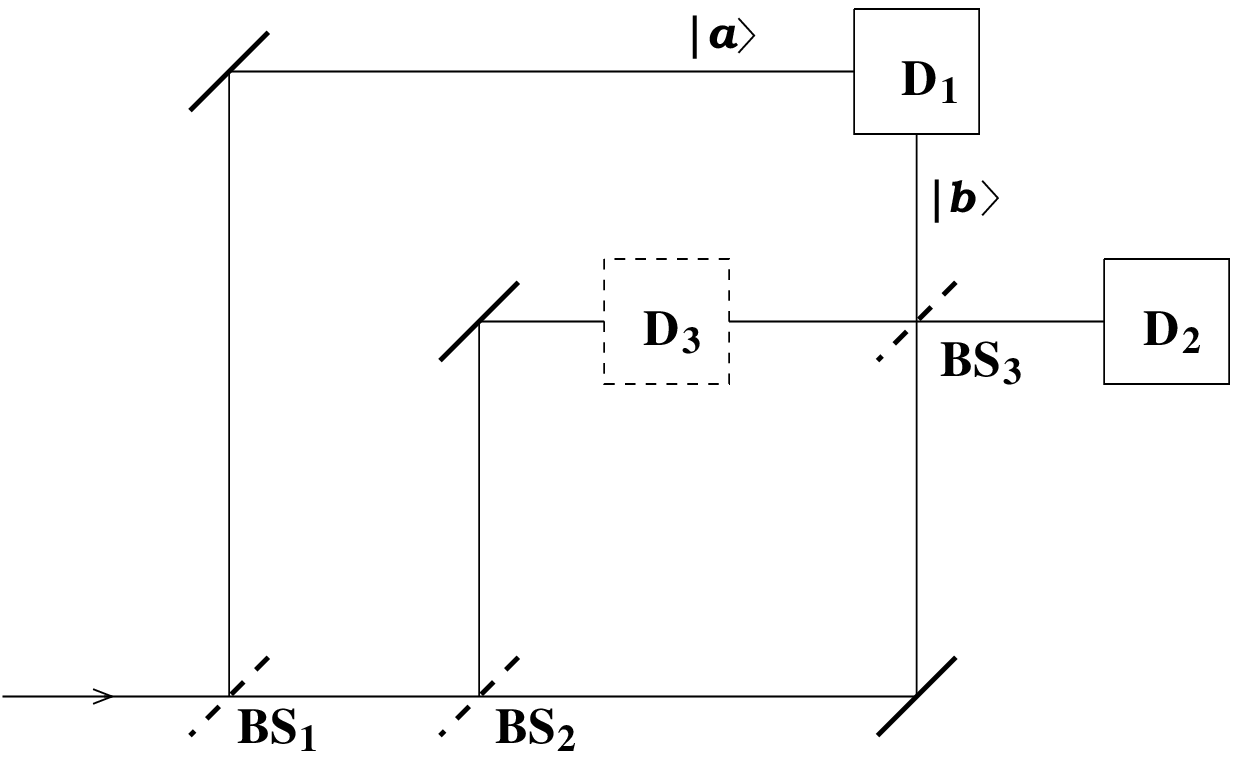}}
\vskip .2cm
\noindent
{\bf Fig. 1 Cohen's experiment.}  Mach-Zehnder type interferometer
  with ``which way'' detector $D_3$ in place.
 
The argument of Cohen is as follows.  If detector $D_3$ is not placed,
then the probabilities for getting the click in $D_1$ and in $D_2$ are
equal, ${\rm Prob}(D_1)={\rm Prob}(D_2) ={1\over 2}$. This is so
because 
the beam-splitters are half-transparent and the Mach-Zehnder $BS_2$ -
$BS_3$ is tuned in such a way that all particles moving towards $BS_2$ are detected
by $D_2$. If $D_3$ is in place, then the ABL formula
yields (Cohen claims) a probability $1\over 4$ for a click in $D_3$ given
a click in $D_1$, ${\rm Prob}(D_3|D_1) = {1\over 4}$, and a probability $1\over2$ for a click in $D_3$ given
a click in $D_2$,  ${\rm Prob}(D_3|D_2) = {1\over 2}$.  Combining these three statements Cohen obtains
an unconditional probability for the click in $D_3$:
\begin{eqnarray}
{\rm Prob}(D_3) =  \nonumber
{\rm Prob}(D_3|D_1) {\rm Prob}(D_1) + ~~~~~~~~~~~~\\
 ~~~~~~~~~{\rm Prob}(D_3|D_2)  {\rm Prob}(D_2)
 = {1\over 4} {1\over 2} +  {1\over 2}
{1\over 2} = {3\over 8} .
\end{eqnarray}
This result  is in contradiction with   quantum theory which
yields ${\rm Prob}(D_3) = {1\over 4}$.

One difficulty with this derivation is that the versions of the ABL
formula which were published so far are not applicable to  Cohen's
experiment. The original ABL formula is applicable to a situation in
which there is a complete measurement at $t_1$, a complete measurement
at $t_2$ and a complete measurement at time $t$, $t_1<t<t_2$.
``Complete'' means that the outcome specifies the quantum state
completely. In Ref.\cite{AV91} a generalization of the ABL formula
which is applicable  for
an arbitrary
measurement at time $t$ is given, but the measurements at $t_1$ and
$t_2$ have to be complete.\cite{foot} 
 The ABL formula of  Ref.\cite{AV91} yields the
probability for the result
$C=c_n$  at time $t$  given that at time $t_1$  the system was prepared in the  state
 $|\Psi_1\rangle$ and that at time $t_2$  the state
$|\Psi_2\rangle$ was found (here, for simplicity,  zero free Hamiltonian is assumed):
  \begin{equation}
  \label{ABL}
 {\rm Prob}(C=c_n) = {{|\langle \Psi_2 | {\bf P}_{C=c_n} | \Psi_1 \rangle |^2}
\over{\sum_i|\langle \Psi_2 | {\bf P}_{C=c_i} | \Psi_1 \rangle |^2}} .
\end{equation}
In  Cohen's example, however, the
measurement at time $t_2$ is not complete: the click in $D_1$ does not
distinguish between  $|a\rangle$, the state of the particle arriving from the left
and $|b\rangle$, the state of the particle arriving vertically from beam-splitter
$BS_3$.  I will analyze the proper generalization of the ABL formula
for this case below and I will reach a different result for
the conditional probability ${\rm Prob}(D_3|D_1)$ which,
nevertheless, will not change Cohen's argument. However, I believe  that
for trying to show putative 
inconsistency of the ABL formula it is better to  consider situations in
which the present version of this formula (shown in Eq. 2) is applicable. Therefore, I will  first
modify Cohen's experiment in such a way that Eq. 2  is
applicable while Cohen's argument  still goes through.
 
In the simplest variation  of the experiment which makes the final
measurement complete,  $D_1$ is modified in such a way that it 
distinguishes the particles in  states $|a\rangle$ and
$|b\rangle$. This, however,  is not suitable for our purpose since it will not
lead to  Cohen's type contradiction.  Therefore, we will
consider, instead, a detector $D_1$ which  distinguishes between
the states 
$|+\rangle \equiv {1\over \sqrt 2} (|a\rangle + |b\rangle)$ and $|-\rangle \equiv {1\over \sqrt 2} (|
a\rangle - |b\rangle)$.
For such an experiment the ABL formula (2) is  applicable
directly and it yields 
${\rm Prob}(D_3|D_{1},+)={1\over 10}$ and
${\rm Prob}(D_3|D_1,-)={1\over 2}$.  If detector $D_3$ is not placed,
then $ \rm{Prob} (D_{1},+) =\rm{Prob} (D_{1},-) = {1\over
  4}$. The probability for detection by $D_2$ remains unchanged: ${\rm
  Prob}(D_2)= {1\over 2}$.
Following Cohen's proof we combine the above results and obtain
the unconditional probability for a click in $D_3$:
\begin{eqnarray}
{\rm Prob}(D_3) \nonumber =  \nonumber {\rm Prob}(D_3|D_{1},+) {\rm Prob}(D_1,+) +~~~\\
\nonumber {\rm Prob}(D_3|D_{1},-) {\rm Prob}(D_1,-) + ~~\\
\nonumber {\rm Prob}(D_3|D_2)  {\rm Prob}(D_2) =~  \\
{1\over 10}{1\over 4}  + {1\over 2} {1\over 4}
+ {1\over 2}
{1\over 2}
 = {2\over 5}.
\end{eqnarray}
Again this differs from  the prediction of quantum theory:  ${\rm Prob}(D_3) = {1\over 4}$.

 Equation (3) is indeed wrong, but not because the probabilities
given by the ABL formula are incorrect. The probabilities ${\rm
  Prob}(D_1,+)$, ${\rm Prob}(D_1,-)$ and ${\rm Prob}(D_2)$ are
obviously wrong. Indeed, these probabilities were calculated on the
assumption that detector $D_3$ was not placed. But equation (3) yields
the probability for click in $D_3$. Therefore, it must be in place
and the assumption on which the probabilities ${\rm Prob}(D_1,+)$,
${\rm Prob}(D_1,-)$ and ${\rm Prob}(D_2)$ were calculated is not
fulfilled.

It is easy to correct the calculation of ${\rm Prob}(D_3)$.
If $D_3$ is in place we obtain:
${\rm Prob}(D_1,+) = {5\over 8}$, ${\rm Prob}(D_1,-) = {1\over 8}$ and
${\rm Prob}(D_2) = {1\over 4}$. Therefore, the correct calculation is:
\begin{eqnarray}
{\rm Prob}(D_3) \nonumber =  \nonumber {\rm Prob}(D_3|D_{1},+) {\rm Prob}(D_1,+)+~~~\\
\nonumber {\rm Prob}(D_3|D_{1},-) {\rm Prob}(D_1,-) + ~~\\
\nonumber {\rm Prob}(D_3|D_2) {\rm Prob}(D_2)  =~ \\
 {1\over 10}{5\over 8} + {1\over 2} {1\over 8}
+{1\over 2} {1\over 4}
 = {1\over 4}.
\end{eqnarray}
Not surprisingly, it is the same number which can be immediately
obtained using quantum rules without considering the results of the
final measurement.

Let us now analyze  the unmodified Cohen's experiment. To this end we
have to first generalize the ABL formula (2) for the case of an incomplete
final measurement.
The proof of Eq. (2)  is given on p. 2317 of Ref.\cite{AV91} and it can
be repeated line by line for the case of incomplete final measurement,
say with the result $B=b$. Essentially,  the only change required is
the replacement of  ``$|\Psi_f\rangle =
|\Psi_2\rangle$''  by ``$B=b$'' and the final result will be:
 \begin{equation}
  \label{ABL-new}
 {\rm Prob}(C=c_n) = {{ || {\bf P}_{B=b} {\bf P}_{C=c_n} | \Psi_1 \rangle ||^2}
\over{\sum_i||  {\bf P}_{B=b} {\bf P}_{C=c_i} | \Psi_1 \rangle ||^2}} .
\end{equation}

Now we can analyze  Cohen's original example. 
The  generalized ABL formula (5) yields: ${\rm Prob}(D_3|D_1) = {1\over 6}$ (instead of
$1\over 4$ in Cohen's paper),
and ${\rm Prob}(D_3|D_2) = {1\over 2}$. 
If we calculate ${\rm Prob}(D_3)$ using the probabilities for the clicks
in $D_1$ and $D_2$ calculated on the assumption that $D_3$ is absent,
Cohen's contradiction  still holds:
\begin{eqnarray}
{\rm Prob}(D_3) =  \nonumber
{\rm Prob}(D_3|D_1) {\rm Prob}(D_1) + ~~~~~~~~~~~~~~~~~~\\
 ~~~~~~~~~{\rm Prob}(D_3|D_2)  {\rm Prob}(D_2)
 = {1\over 6} {1\over 2} +  {1\over 2}
{1\over 2} = {1\over 3}~ (\neq {1\over4})
\end{eqnarray}
But  the correct calculation of probabilities of the detection by
$D_1$ and $D_2$ (which assumes that detector
$D_3$ is present) 
yields: ${\rm Prob}(D_1) = {3\over 4}$ and ${\rm Prob}(D_2) = {1\over 4}$. Thus, we
obtain again the correct result:
\begin{eqnarray}
{\rm Prob}(D_3) =  \nonumber
{\rm Prob}(D_3|D_1) {\rm Prob}(D_1) + ~~~~~~~~~~~~\\
 ~~~~~~~~~{\rm Prob}(D_3|D_2) {\rm Prob}(D_2) 
 = {1\over 6}{3\over 4} + {1\over 2} {1\over 4}
 = {1\over 4}.
\end{eqnarray}

Since Cohen's  proof of inconsistency of the ABL rule is erroneous, his
conclusions about the  limitations on the applicability of the ABL
rule are unfounded. The surprising consequences of the ABL rule in the
examples which Cohen considered are still valid.

This research was supported in
part by a grant No. 614/95 from the Israel Science  Foundation.

\end{multicols}


\begin{thebibliography}{9}
\bibitem{foot}
In Ref.\cite{AV91} there is also another generalization of the ABL formula
for  systems described by ``generalized states'' which are defined
there. However, this generalization is not relevant for the present problem.
\end{thebibliography}
\end{document}